# Enhanced self-attraction of proteins and its evolutionary implications.

D. B. Lukatsky and E. I. Shakhnovich
*Department of Chemistry and Chemical Biology, Harvard University, Cambridge MA 02138*

**Abstract:**
Statistical analysis of protein-protein interactions shows anomalously high frequency of homodimers [Ispolatov, I., *et al.* (2005) *Nucleic Acids Res* **33**, 3629-35]**.** Furthermore, recent findings [Wright, C.F., *et al.* (2005) *Nature* **438**, 878-81] demonstrate that maintaining low sequence identity is a key evolutionary mechanism that inhibits protein aggregation. Here, we study statistical properties of interacting protein-like surfaces and predict the effect of universal, enhanced self-attraction of proteins. The effect originates in the fact that a pattern self-match between two *identical*, even randomly organized interacting protein surfaces is always stronger compared to the pattern match between two *different*, promiscuous protein surfaces. This finding implies an increased probability of homodimer selection in the course of early evolution. Our simple model of early evolutionary selection of interacting proteins accurately reproduces the experimental data on homodimer interface aminoacid compositions. In addition, we predict that heterodimers evolved from homodimers with the negative design evolutionary pressure applied against promiscuous homodimer formation. We predict that the anti-homodimer negative design evolutionary signal is conveyed through the enrichment of heterodimeric interfaces in polar residues, and most profoundly in glutamic acid and lysine, which is consistent with experimental findings. We predict therefore that the negative design against homodimers is the origin of the observed, highly conserved polar "hot spots" on heterodimeric interfaces.

**Introduction**

Significant effort has been devoted to the studies of protein-protein interactions (PPI) and a number of interesting observations emerged. In particular it was shown recently that homodimers occur with anomalously high frequency(1-3). Recent analysis of PPI networks of four eukaryotic organisms (baker's yeast *S.cerevisiae*, nematode worm *C.elegans*, the fruitfly *D.melanogaster* and human *H.sapiens*) obtained from high-throughput experiments reported that the actual number of homodimeric proteins is 25-200 times higher than expected if such homodimers randomly appeared in the course of evolution(1). Further, universal preference for homodimeric interactions (a phenomenon called "molecular narcissism") is apparent in detailed analysis of confirmed protein-protein interactions (S. Teichman, private communication). It was also shown experimentally(4) that the diversity of protein sequences is a major factor in reducing the propensity of proteins to aggregate. These striking observations remain unexplained. Here, we propose a simple model of protein-protein interactions and show that observed preference for homodimeric complexes is a consequence of general property of protein-like interfaces to have high affinity for self-attraction, as compared with propensity for attraction between *different* proteins. In particular, we noticed that even for random protein-like interfaces the self-attraction is always statistically stronger compared with promiscuous interactions between different random interfaces. Our analysis suggests a simple evolutionary one-shot scenario with the enhanced probability for the emergence of homodimeric complexes and provides guidance of how subsequent evolution of selective heterodimeric complexes proceeded.

**Enhanced self-attraction of model protein surfaces**

We use a simple, residue-based model of a protein interface(5) (see Figure 1) where amino acids are represented as hard spheres of diameter $\sigma = 5 \text{Å}$ confined to a planar, circular interface of diameter $D = 70 \text{ Å}$. Each model interface is built by randomly placing residues of all twenty aminoacid types, on the surface and by fixing the obtained configuration, Figure 1. The aminoacid compositions are specified by the probability distribution, and thus the compositions of different interacting surfaces (IS) vary, but the total number of residues, $N$, in each interface is fixed, $N = 70$. The surface fraction $\rho$ of residues on an interface (the reduced surface density) is $\rho = N\sigma^2/D^2 \simeq 0.357$. The chosen parameters correspond to a typical protein interface(3, 6, 7). Residues of two IS interact via the Miyazawa-Jernigan (MJ) residue-residue potentials(8), and we assume that two residues are in contact if they are separated by the distance less than $8 \text{ Å}$.

We investigated the statistical interaction properties of such IS at various random realizations of aminoacid placements on IS. For each realization of surfaces we fixed the inter-protein separation to be $5.01 \text{ Å}$, and rotated each pair of superimposed surfaces to find extreme, lowest value of interaction energy for this pair. This way we obtained the extreme value distribution (EVD) of the inter-protein interaction energies, $E$, between different random realizations of IS in two cases: (i) random heterodimers (superimposed pairs of *different*, random surfaces) and (ii) homodimers (mirror-image self-superimposed surfaces). The results of these calculations for different, *average* aminoacid compositions are shown in Figure 2. The key result is that random model protein interfaces have *always* a statistically higher propensity for self-attraction as compared with random heterodimers. The tail of the EVD for homodimers is *always* shifted towards lower energies with respect to random heterodimers. That means that it is significantly more probable to find strong homodimeric complexes in random "soup" of protein interfaces than it would be expected if such complexes were selected at random (*i.e.* simply selected based on their average concentrations). The predicted effect of enhanced self-attraction is universal and has very simple physical explanation as follows. Although locations and identities of residues on each surface are random and disordered, two *identical*, random surfaces are *always* more likely to strongly attract each other, as compared to two different random surfaces because it is always easier to match a random pattern with itself (an automatic match) than with another random pattern (a much less likely event). Figure 3 demonstrates the origin of this effect. We computed the number of inter-surface, residue-residue contacts, *n*, for each case represented in Figure 2, and constructed the corresponding probability distributions, *P(n)*, for random heterodimers and homodimers, respectively (see Figure 3). The key observation here is that the right tails of homodimer *P(n)* are *always* shifted towards the higher number of contacts as compared with heterodimer *P(n)*. The universally enhanced structural similarity of self-interacting surfaces (even random surfaces) leads to the higher maximal number of favourable, inter-surface contacts, which in turn, enhances the self-attraction of surfaces. The phenomenon of the enhanced self-attraction of protein interfaces represents the central finding of this paper. We emphasize that the strength of the effect depends on the composition of interfaces, however the effect itself is universal and holds for *any* composition (Figure 2, black curves). We also stress that the predicted effect is statistical in its nature, and holds universally for protein sets, rather than for individual proteins: It is not necessarily that *every* homodimer has a lower interaction energy than *any* heterodimer, but rather the probability distribution of interaction energies, *P(E)*, for homodimers is necessarily shifted towards lower energies as compared with *P(E)* for heterodimers.

**One-shot evolutionary selection**

Our results imply that homodimers could have been selected with higher probability (than would be expected randomly) in the course of prebiotic evolution as first functional protein-network motifs as a result of a possible "one-shot selection" of strongly interacting proteins from the pool of proteins exposing random surfaces. In order to check whether such scenario indeed took place we simulated one-shot selection by simply selecting strongly self-interacting surfaces (*e.g.* with energy of interaction *E*<-3.3, Figure 2 (*e*)) from the pool of all randomly generated ones (left tail of homodimer *P(E)* in Figure 2 (*e*)). Next we checked aminoacid composition in such "one-shot" selected, strongly interacting homodimeric surfaces and compared it with observed compositions in homodimeric interfaces of proteins(6, 7). The resulting aminoacid composition of one-shot *selected*, strongly attracting homodimeric IS is presented in Figure 4 (*b*) in terms of the interface propensity for each of 20 residues. The model interface propensity is $\ln(f_\alpha / f_\alpha^0)$, with $f_\alpha$ and $f_\alpha^0$ being the fraction of residue type $\alpha$ in the *selected* set of surfaces and the average fraction of residue $\alpha$, on *all* protein surfaces (which coincides with probability distribution with which we selected aminoacid types to generate random surfaces as explained above), respectively. We emphasize that $f_\alpha^0$ is the input to the model from experimental data, and $f_\alpha$ is produced by the model. The model results correlate with the observed experimental interface propensities(7) with the correlation coefficient $R \simeq 0.93$, Figure 4 (*c*) (see also Methods). Such strong correlation between model and experiment suggests an evolutionary scenario of initial "one-shot", preferential selection of strongly interacting, natural homodimers. Indeed if we change selection criterion from that of low-energy self-interacting surfaces to a "window" of higher interaction energies we observe a sharp transition in aminoacid composition of surfaces selected in a sliding window of interaction energies, from the highly correlated with experiment value of +0.93 (when strongly interacting surfaces corresponding to the left tail of the EVD on Figure 5 are selected) to the anti-correlated with experiment value of –0.91 (when mutually repulsive homodimeric surfaces at the right tale of the EVD are selected), see the legend of Figure 5. We emphasize that such high correlation between the model predictions and experimental data is much more than just a correct yet trivial prediction for the relative propensity of hydrophobic and hydrophilic residues at protein homodimeric interfaces. The model correctly predicts the relative propensities *within* the hydrophobic and hydrophilic groups of amino acids. This is demonstrated in the reshuffling control calculation presented in Supporting Information. Finally, we stress that the predicted high correlation between the model and experiment is robust with respect to the choice of the effective, residue-residue potentials (see Supporting Information).

**Homodimers-to-heterodimers paralogous divergence and negative design**

We now turn to the question of how homodimers evolved after the predicted "one-shot" evolutionary selection. To answer this question we performed the stochastic design procedure (see Methods) to mimic the evolutionary transformation of homodimers towards heterodimers. Our aim is to compare the resulting model and experimental difference between the homodimeric(7) and heterodimeric(9) interface propensities. The stochastic design algorithm started from strongly interacting homodimeric interfaces and proceeded to evolve them to strongly interacting heterodimeric interfaces by duplication and paralogous divergence. In particular, we duplicated each selected interface and evolved it to heterodimer by random mutations (of each interacting surface within the interacting pair) with the Metropolis acceptance criterion that minimizes the inter-protein interaction energy (see Methods). In addition to the requirement of strong interaction between surfaces we also applied a negative design requirement against promiscuous homodimer formation. We also evolved independently all initial "seed" homodimeric interfaces as homodimers, by simply minimizing their interaction energy. At the end of this procedure we compared the resulting compositions of homodimeric and heterodimeric interfaces. The difference between the homodimeric and heterodimeric propensities is shown as a scatter plot in Figure 6. Our key observation here is that the heterodimeric design results in enrichment of the interface with polar residues and especially significantly with Glu and Lys. Application of the negative design pressure against homodimers strengthens this effect – the higher is the strength of the applied negative design, the richer is the interface in Glu and Lys (see Figure 7). This is in agreement with the experimental data on heterodimeric interfaces, Figure 6, and with other investigations of the effect of negative design(10, 11). We emphasize that the stochastic design procedure results in highly non-random placement of Glu and Lys residues within the interfaces, making specific salt bridges, providing highly specific and strongest effect of negative design against homodimers. This is apparent already in Figure 2 (*d*), where the extreme case of Glu-Lys-rich (and random) interfaces is shown. This observation demonstrates that only compositional (with entirely random *positional* distribution) enrichment with Lys and Glu is not sufficient to shift the low-energy tail of the homodimer EVD to higher energies with respect to the tail of the promiscuous heterodimer EVD.

The overall linear correlation coefficient $R$ between predicted and observed heterodimer interface compositions is not very high, $R \simeq 0.57$ (Figure 6 (*b*)). This is due to three outlier residues in the scatter plot: Trp, Tyr and Ala. Removing these three residues from the scatter plot would improve $R$ significantly, to the highly correlated value of $R \simeq 0.87$. The existence of the outliers indicates that our design procedure does not capture some additional evolutionary pressures that were applied to evolving (diverging) heterodimers. Trp has the most pronounced deviation from the linear plot in Figure 6. It has been noticed in Ref.(12), that Trp is highly conserved at the interface of heterodimers. This strongly indicates the existence of an additional yet unexplained evolutionary pressure that leads to enrichment of heterodimer interfaces with tryptophan residues.

**Conclusion**

In summary, our findings provide rationale for many recent statistical observations of protein-protein interactions and provide plausible scenario for their evolution. We show that homodimer formation is statistically more likely than having occurred fully by chance with the same probability as heterodimers. Further we show that a plausible mechanism of some heterodimer formation from initially selected homodimers. That is not to say that heterodimers could not have evolved in early evolution in a similar one-shot selection mechanism. In fact while heterodimers indeed have statistically lower propensities to interact they may be favoured in some cases entropically due to greater diversity of possible heterodimeric interfaces. Nevertheless we show here that peculiar universal self-pattern matching phenomenon shifts distribution of protein-protein interfaces towards greater proportion of homodimers, in harmony with experimental observations. A key support to this view comes from comparison of aminoacid interface propensities in real and model homodimeric interfaces shown in Fig.4. Extremely significant and robust with respect to the choice of potential correlation of R=0.93 supports the ''one-shot'' selection mechanisms where homodimers were selected early in evolution as strongly self-interacting pairs in the set of randomly exposed protein surfaces. Further, our findings explain the recent experimental discovery(4) that the diversity of sequence identity is a major evolutionary mechanism inhibiting protein aggregation and amyloid formation. It was demonstrated in(4) that in two large, multidomain protein superfamilies (immunoglobulin and fibronectin type III) of the adjacent domain pairs in the same proteins, more than two-thirds have less than 30% identity. Moreover, only about 10% of the adjacent domain pairs have more than 40% identity. It is the key message in(4) that the low sequence similarity protects domains from unwanted aggregation. Our prediction of universally enhanced self-attraction of proteins rationalizes the anti-aggregation protection mechanism discovered by Dobson *et al.*(4) as evolutionary emerged from the statistically enhanced correlations between *identical* and even disordered protein interfaces as compared with *different*, promiscuous interfaces.

**Acknowledgements.** We are grateful to S. Teichman for communicating to us her unpublished results. This work is supported by NIH grant GM52126.

**Methods**

**Aminoacid interface propensities, model vs. experiment**

The experimental interface propensity of a given residue type $\alpha$, is defined as a logarithm of the ratio, $\ln(f_\alpha / f_\alpha^0)$, where $f_\alpha$ is the protein *interface* (*i.e.* the part of a protein surface that takes part in the interaction with another protein) area fraction (also called composition) of residue type $\alpha$, and $f_\alpha^0$ is the total (*i.e.* over the entire surface) protein *surface* fraction of this residue type(6, 7), see Figure 4 (*a*). Both $f_\alpha$ and $f_\alpha^0$ are obtained as averages with respect to a set of protein complexes(6, 7). The corresponding model aminoacid interface propensities are defined analogously, where $f_\alpha^0$ are the *average* compositions used to generate the entire random set of surfaces (*e.g.* these average compositions are indicated in each plot of Figure 2). The model $f_\alpha$ are the resulting compositions obtained from the compositional analysis of the *selected* set of interacting surfaces (out of all generated surfaces), using selection criteria described in the main text.

**Stochastic design procedure**

The stochastic design procedure attempts a mutation by randomly changing the identity of a randomly chosen residue within each of the two interacting surfaces. The attempted mutation is accepted with the standard Metropolis criterion(13) on the extreme value of the inter-protein interaction energy. The extreme value of the inter-protein interaction energy is computed in each MC step. The negative design on homodimer formation is implemented in the MC procedure using the total inter-protein energy in the form $E_{tot} = E_{hetero} - \alpha E_{homo}$, where $E_{hetero}$ and $E_{homo}$ are the interaction energy of the corresponding hetero- and homodimer pair, respectively, and the strength of the negative design $\alpha$ is chosen to be 0.5 in computing Figure 6. In all MC design runs $2N=140$ mutation attempts for each protein surface (within each pair of superimposed surfaces) were performed. The effective, design temperature, $T$, entering the Boltzmann factor of the Metropolis criterion(13), $\exp(-E_{tot}/T)$, was chosen to be $T=4$. All simulations were performed starting with 20000 random protein interfaces.

**Figure Legends:**

**Figure 1:** Snapshot of a typical, random configuration of a model protein interface.

**Figure 2:** Computed extreme value distribution (EVD), $P(E)$, of the interaction energy, $E$, between two model protein interfaces for random heterodimers (red line) and homodimers (black line). $E$ is the interaction energy per one residue in the units of $k_B T$, where $k_B$ is the Boltzmann constant. The aminoacid composition was chosen to be uniform in (*a*), *i.e.* all residue types have equal, 5% probability of occurrence; *E.coli* composition in (*b*); 50% Ala and 50% His in (*c*); 50% Lys and 50% Glu in (*d*); and the composition of homodimer surfaces data set of Table III in Bahadur *et al.*(7) in (*e*).

**Figure 3:** Computed probability distribution, *P(n)*, of the total number of inter-surface residue-residue contacts, *n*, per residue between two model protein surfaces, at the extreme value of the inter-surface energy, *E*. For each pair of interacting surfaces the extreme, lowest value of inter-surface interaction energy is found (exactly in the way described in the legend of Figure 2), and *n* is computed in the mutual surface orientation corresponding to this extreme value of *E*. The colour code is exactly the same as in Figure 2: random heterodimers (red line) and homodimers (black line). The average aminoacid compositions in each plot are also in exact correspondence with Figure 2. The gaps in *P(n)* originate in the fact that the number of inter-surface residue-residue of contacts can vary only by an *integer* number, at least by one. The right homodimer tails of *P(n)* are always shifted towards the *higher* number of contacts as compared with the right tails of random heterodimers.

**Figure 4:** Comparison of experimental data on the compositions of protein interfaces and model predictions. (*a*): Compositions of homodimeric (Ref.(7), Table III) and heterodimeric (Ref.(9), Table 4 and Ref.(14), Table 2) protein interfaces in terms of interface propensities, $\ln(f_\alpha / f_\alpha^0)$, where $f_\alpha$ is the protein *interface* area fraction of residue type $\alpha$, and $f_\alpha^0$ is the total protein *surface* area fraction of residue type $\alpha$. (*b*): Computed propensities of homodimeric interfaces based on the evolutionary selection model (selection of only those of random interfaces that form low energy homodimers). The average compositions of residues used to generate random interfaces are taken from the homodimer data set of Bahadur *et al.*(7) (Table III, column 5 (Surface) of Ref. (7), with surface compositions for homodimers in terms of area fraction; these compositions were used also to compute **Figure 2** (*e*)). The inset in (*b*) shows the corresponding EVD (identical to **Figure 2** (*e*)) and the arrow indicates the cut-off value of energy, $E = -3.3$, below which the selection of homodimers is performed (the line colour code is identical to the one used in **Figure 2**). (*c*): The scatter plot of experimental vs. model residue interface propensities for homodimers. The resulting linear correlation coefficient between the experimental and model data is $R \simeq 0.93$. The straight line represents the linear fit to the data.

**Figure 5:** Bottom: Computed correlation coefficient *R* as a function of the position of the energy "window", $(E, E + \Delta E)$, where $\Delta E$ is the width of the energy "window". The resulting dependence *R(E)* is practically independent on the magnitude of $\Delta E$ for small $\Delta E$. Each point on this graph is obtained using the procedure similar to the one used to compute **Figure 4** ((*b*) and (*c*)), and the resulting correlation coefficient *R*, with the only difference being that instead of the selection from the tail of *P(E)*, *E*<3.3, in the present case the selection of homodimers is performed from the sliding energy "window". Sharp transition of *R* corresponds to the enhanced probability, "one-shot selection" of homodimers evolutionary scenario. Top: EVD *P(E)* [identical to **Figure 2** (*e*)] is plotted for the sake of comparison.

**Figure 6:** Experimental (*a*) and computed in the course of the MC design procedure, and represented as a scatter plot (*b*) differences between the interface aminoacid propensities of homodimeric and heterodimeric protein interfaces, respectively (experimental data are from Ref. (7) for homodimers and from Ref. (9) for heterodimers). The negative design on homodimer formation was switched on, as described in the main text and Methods, in computing (*b*). Most of the residue composition differences are predicted correctly except for Trp, Tyr, and Ala. The overall linear correlation coefficient of the data is $R \simeq 0.57$. Excluding these three outliner residues improves *R* to as high as 0.87. The red and blue straight lines in (*b*) represent the linear fit to the entire dataset and the dataset with excluded Trp, Tyr, and Ala residues, respectively.

**Figure 7:** Computed difference between the interface aminoacid propensities for Lys (blue curve) and Glu (red curve) of homodimeric and heterodimeric protein interfaces, respectively, as a function of the strength of the negative design, $\alpha$, in the MC evolution procedure (see Methods and Figure 6 for definitions and explanations). The lower is the value of the computed propensity difference, the richer is the heterodimeric interfaces in Lys and Glu.

**Figure 1**

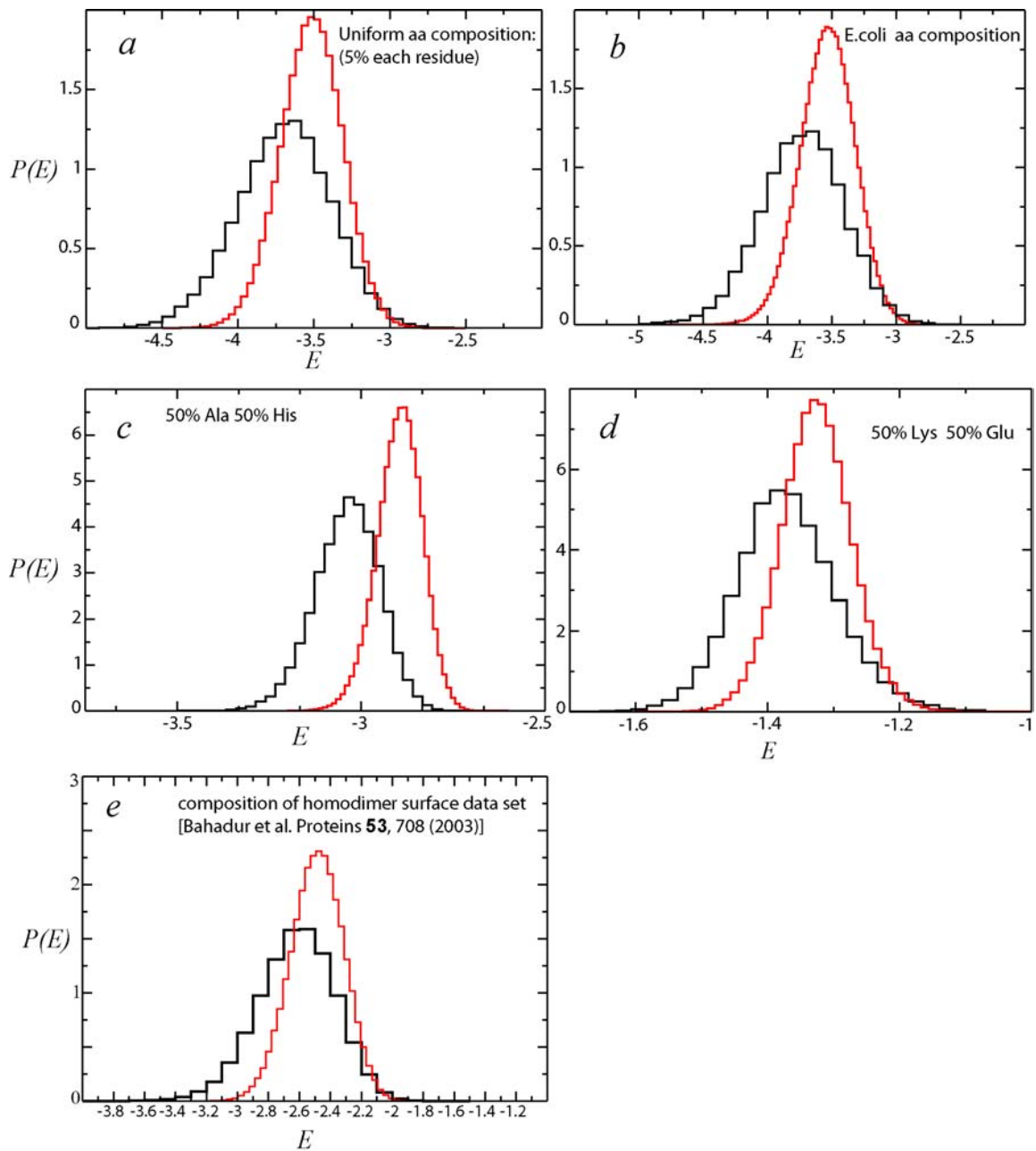

**Figure 2**

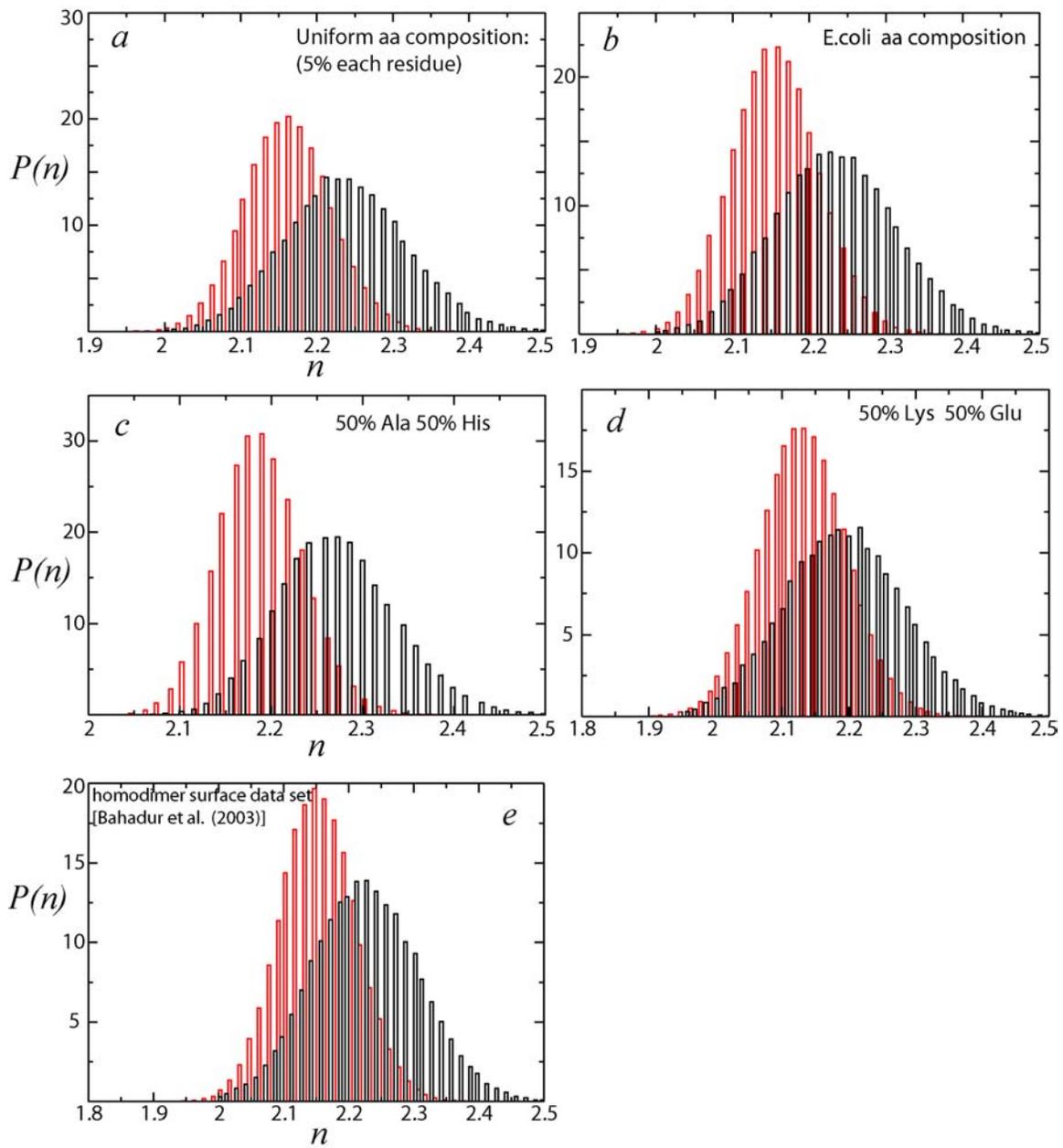

**Figure 3**

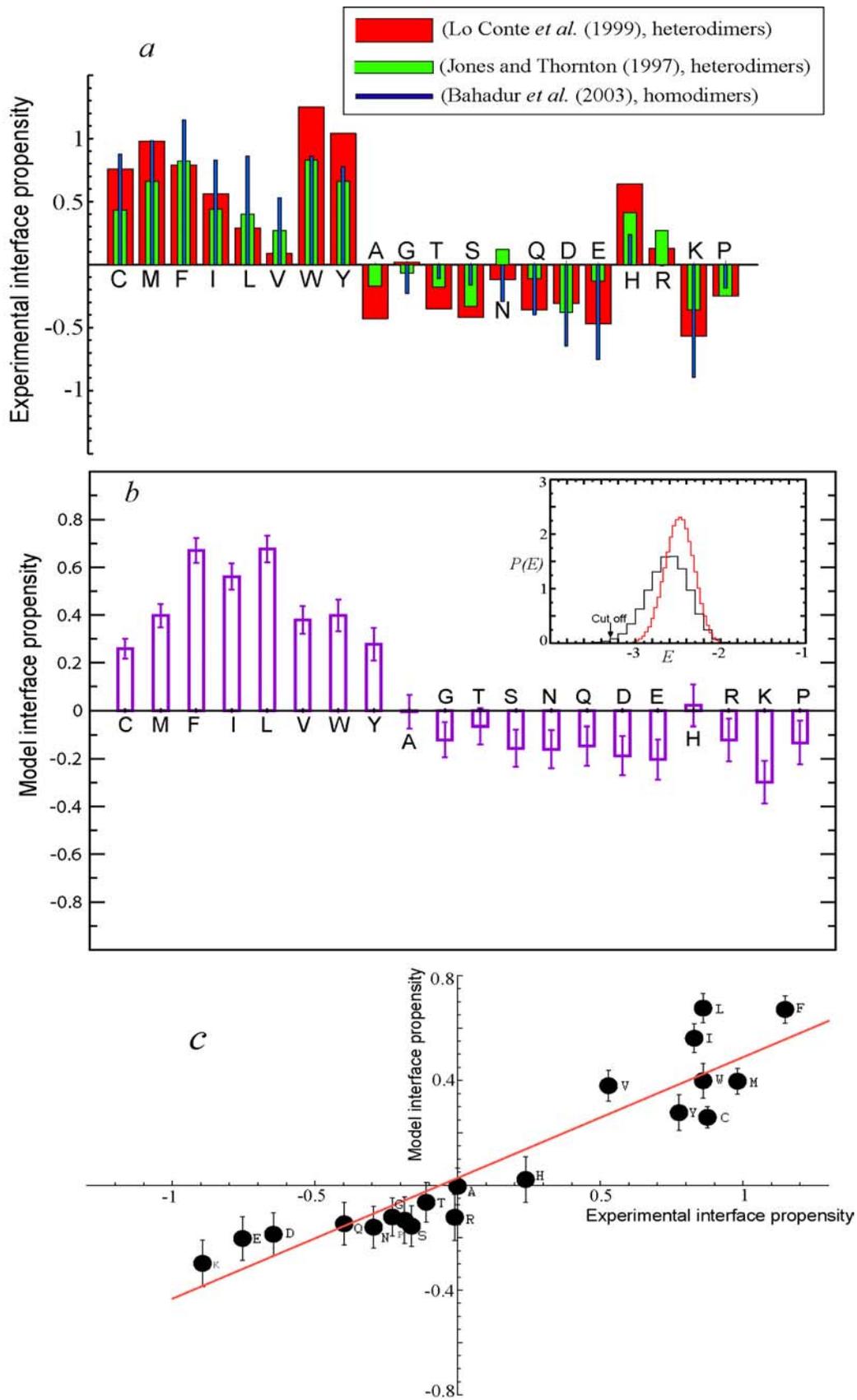

**Figure 4**

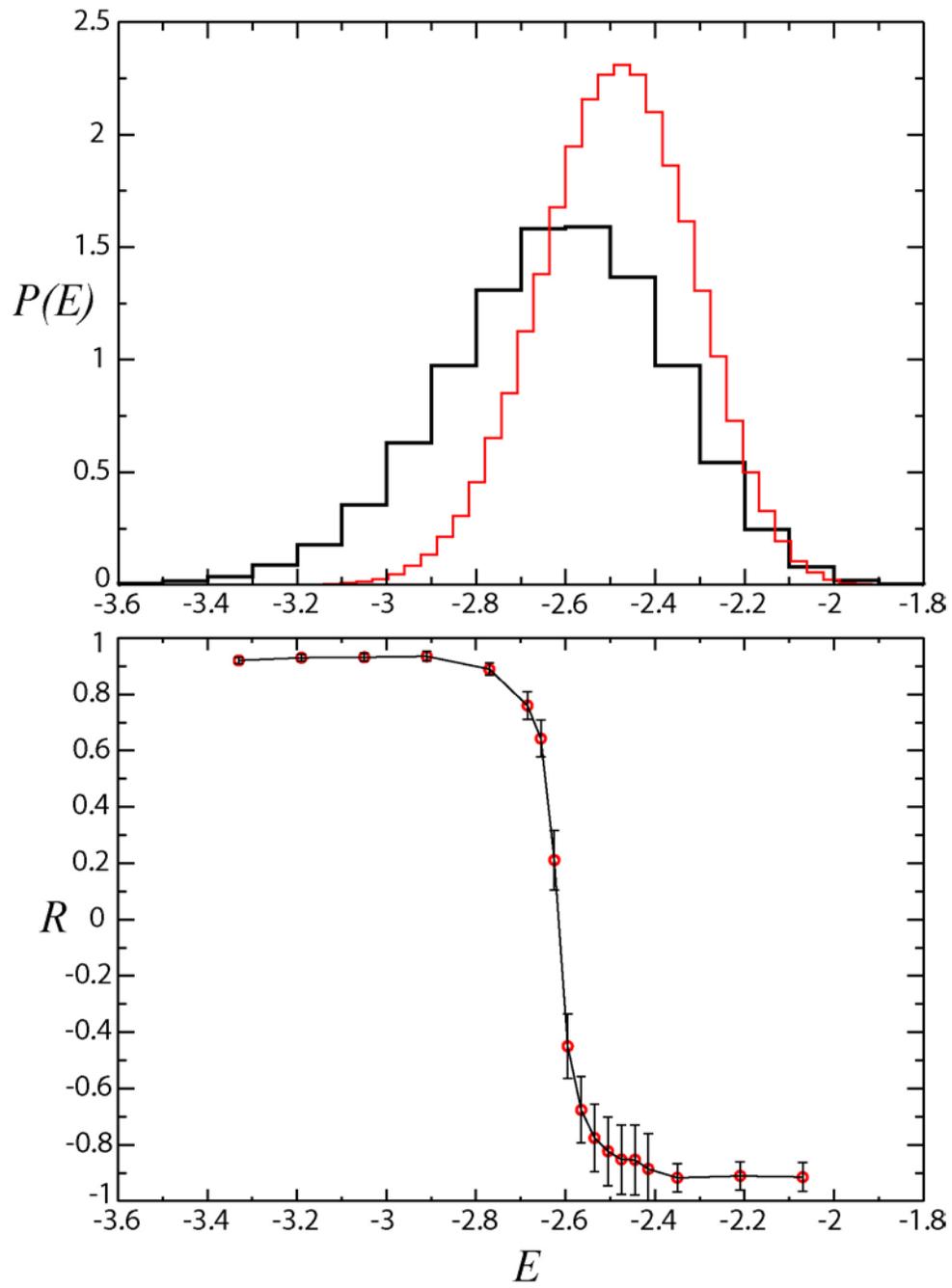

**Figure 5**

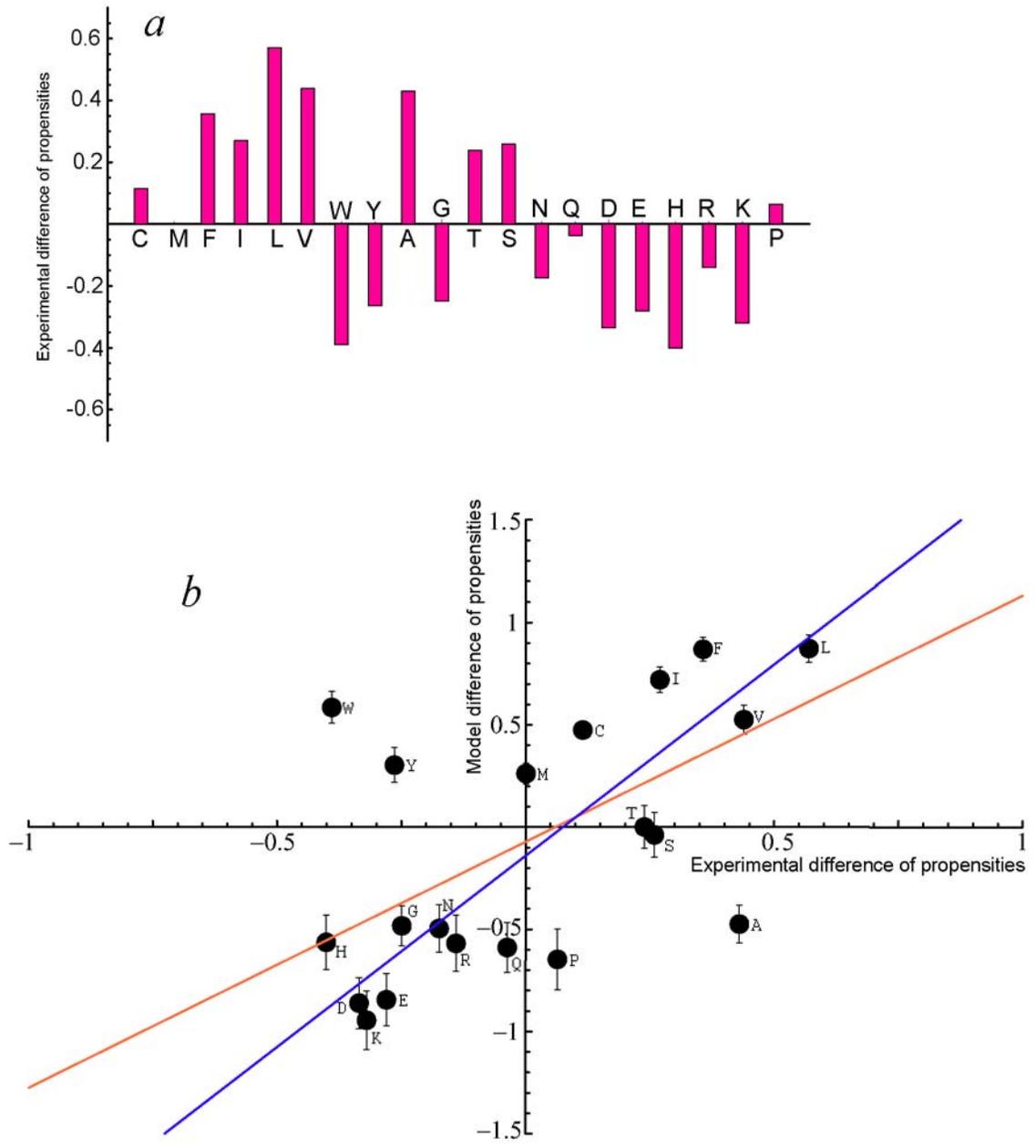

**Figure 6**

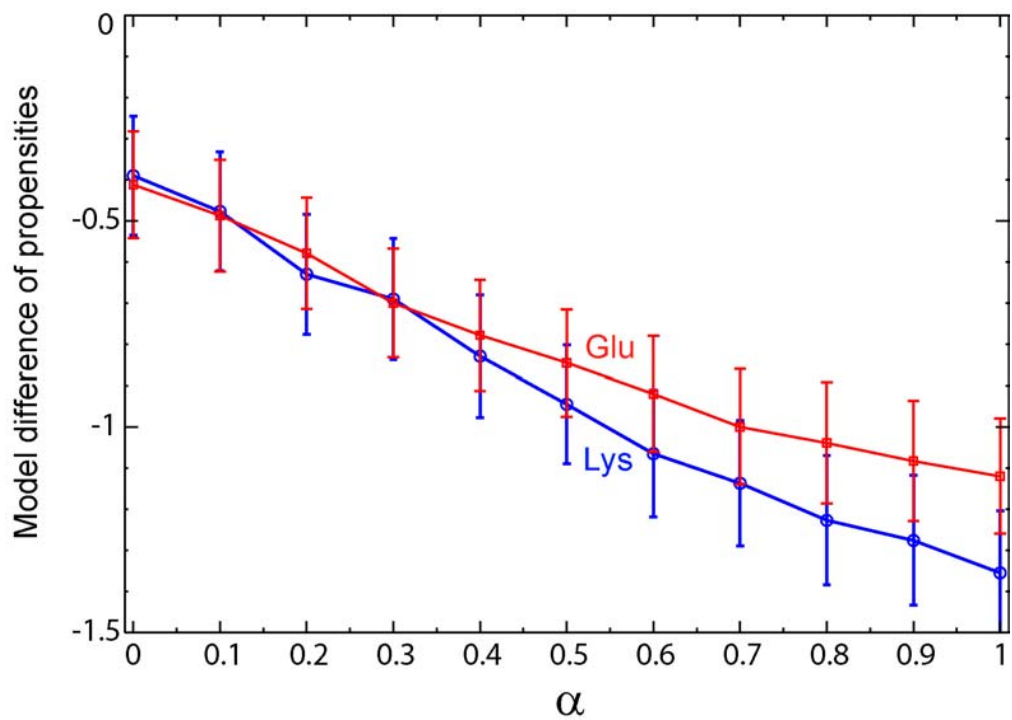

**Figure 7**

**Supporting Information:**

**Reshuffling control**

We computed the probability distribution function of the linear correlation coefficient, $R$, upon the *partial* reshuffling the identities of residues in the model data set, *i.e.* upon reshuffling separately within the mostly hydrophobic [Cys Met Phe Ile Leu Val Trp Tyr Ala] and mostly hydrophilic [Gly Thr Ser Asn Gln Asp Glu His Arg Lys Pro] groups of residues, respectively, Supplementary Figure 1 (*a*). This procedure shows negligibly small probability $p(R>0.93) \simeq 0.00006$ to find the predicted correlation coefficient "by chance" (even assuming a correct redistribution of hydrophobic and hydrophilic residue groups). The complete reshuffling of residue identities shown as a control in Supplementary Figure 1 (*b*) leads, of course, to a symmetrically distributed around zero probability distribution with a zero (up to the computer precision) probability of obtaining $p(R>0.93)$ "by chance".

*One-shot selection with Mirny-Shakhnovich potential*

To verify the robustness of the results on one-shot selection (reported in Figure 4 of the paper) with respect to the choice of the effective, residue-residue interaction potential, we have performed the same calculation using an alternative - the Mirny-Shakhnovich (MS) potential(15) - instead of the MJ potential(8). We followed the procedure identical to the one discussed in detail in the paper. The resulting scatter plot of the model vs. experimental homodimer interface propensities is shown in Supplementary Figure 2 (this plot is the analogue of Figure 4 (*c*)). The linear correlation coefficient $R$ between the model and experimental results is approximately as high, *R=0.91*, in this case, as it was in the case with the MJ potential (*R=0.93* in Figure 4 (*c*)).

**Supplementary Figure 1:** Computed probability distribution *P(R)* of the correlation coefficient *R* upon partial (*a*) and complete (*b*) reshuffling of the residue identities in the model results for the interface propensities. In (*a*) the reshuffling is performed within the two groups of residues: mostly hydrophobic [Cys Met Phe Ile Leu Val Trp Tyr Ala] and mostly hydrophilic [Gly Thr Ser Asn Gln Asp Glu His Arg Lys Pro]. The arrow in (*a*) indicates the position of the predicted value of *R*.

**Supplementary Figure 2:** The scatter plot of experimental vs. model residue interface propensities for homodimers obtained using the MS residue-residue interaction potential(15) instead of the MJ potential(8). The resulting linear correlation coefficient between the experimental and model data is $R \simeq 0.91$. The straight line represents the linear fit to the data.

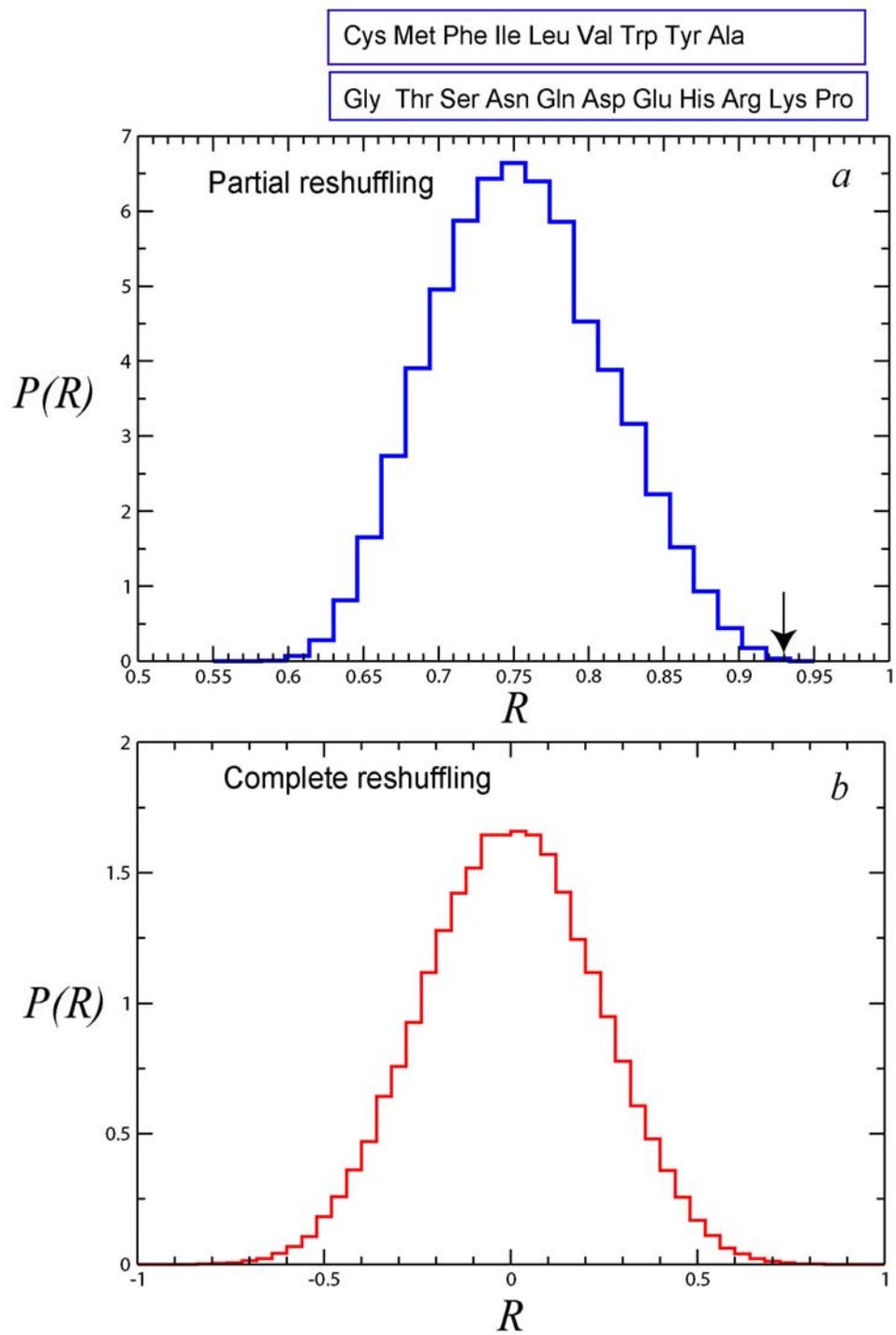

**Supplementary Figure 1**

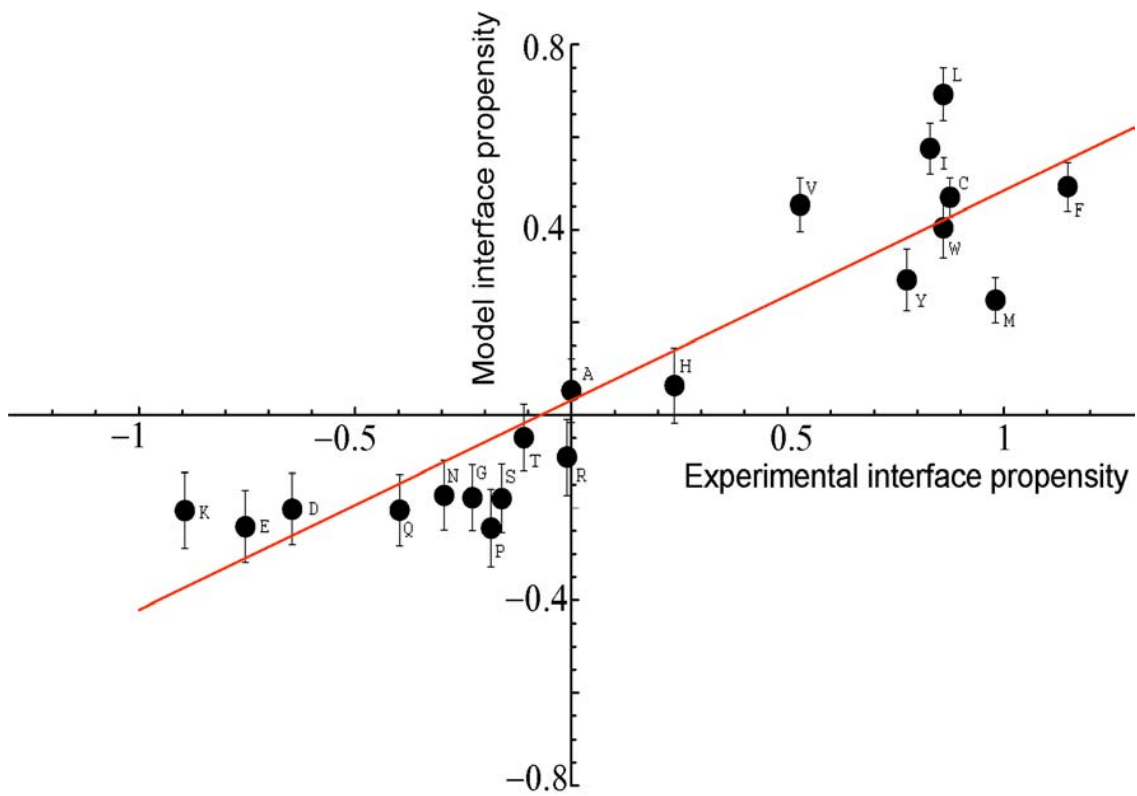

**Supplementary Figure 2**